\documentclass[a4paper]{article}

\usepackage{INTERSPEECH2020}
\usepackage{blindtext} 

\usepackage{makecell}

\title{Mixture of Speaker-type PLDAs for Children's Speech Diarization}
\name{Jiamin Xie$^{1}$, Suzanna Sia$^{1}$, Paola Garc\'ia$^{1,2}$, Daniel Povey$^{3}$, Sanjeev Khudanpur$^{1,2}$}
\address{
  $^1$Center for Language and Speech Processing \& $^2$Human Language Technology Center of Excellence \\ 
  The Johns Hopkins University, Baltimore, MD 21218, USA\\
 $^3$Xiaomi Corp., Beijing, China}

\email{\{jxie27, ssia1, lgarci27\}@jhu.edu, dpovey@gmail.com, khudanpur@jhu.edu}
\begin{document}

\maketitle
%
\begin{abstract}
In diarization, the PLDA is typically used to model an inference structure which assumes the variation in speech segments be induced by various speakers. The speaker variation is then learned from the training data. However, human perception can differentiate speakers by age, gender, among other characteristics. In this paper, we investigate a speaker-type informed model that explicitly captures the known variation of speakers. We explore a mixture of three PLDA models, where each model represents an adult female, male, or child category. The weighting of each model is decided by the prior probability of its respective class, which we study. The evaluation is performed on a subset of the BabyTrain corpus. We examine the expected performance gain using the oracle speaker type labels, which yields an 11.7\% DER reduction. We introduce a novel baby vocalization augmentation technique and then compare the mixture model to the single model. Our experimental result shows an effective 0.9\% DER reduction obtained by adding vocalizations. We discover empirically that a balanced dataset is important to train the mixture PLDA model, which outperforms the single PLDA by 1.3\% using the same training data and achieving a 35.8\% DER. The same setup improves over a standard baseline by 2.8\% DER.

\end{abstract}
\noindent\textbf{Index Terms}: speaker diarization, children's speech, transformer encoder, mixture of PLDAs

\section{Introduction}
Speaker diarization aims to answer the question of "who speaks when?" in a recording. It is the crucial first step that ensures the single-speaker assumption necessary for downstream tasks, including speaker verification and speech recognition, among others.  Most diarization methods take short (1-2 sec) overlapping segments of a recording, estimate similarities between each pair of segments, and cluster/separate segments to same/different speaker(s). The process results in hypothesized speech segments of various lengths that belong to each speaker.

Research in diarization has mainly focused on adult speech. The benchmark diarization error rates (DER) in controlled speech environment, such as telephone conversations or business meetings, typically range from 2\% to 10\% among the best systems \protect\cite{1677976,6135543}. With more natural conditions, such as telephone conversational speech, the diarization performance varies between 5\% and 30\% DER \protect\cite{mccree2019speaker,senoussaoui2013study}. However, these results are made possible under a rather easy setup of few number of speakers or with similar speaking style shared by participants. Recent studies have found diarization is hard under realistic conditions that involve overlapping speech, noisy background, and diverse modalities of speech \protect\cite{watanabe2020chime,garcia2019speaker}. 

Children's speech is one of the realistic domains that poses challenges for speaker diarization \protect\cite{ryant2018first}. The acoustic and linguistic properties of children differ from that of adults, such as higher pitch and formant frequencies and longer phoneme duration \protect\cite{gerosa2006analyzing}. In addition, children utter spontaneous vocalizations during their speech, which increases the need for intra-speaker variations to be appropriately modeled. These spontaneous vocalizations can also occur when others are speaking, which calls for overlap diarization. The above factors introduce a large performance discrepancy between the diarization system of children’s speech and adult speech. One of the studies which analyze the language exposure of children in a home environment revealed on average a 48.9\% DER performance across different training datasets and the state-of-the-art diarization systems \protect\cite{Cristi2018TalkerDI}.

Our previous work \protect\cite{xie2019multi} focused on the adaptation of a PLDA model through both children's speech data augmentation and a discrimination of speaker representations by adult female, adult male, and child type of speakers. In this paper, we extend the later idea to a mixture PLDA model that explicitly captures the variation of speakers across speaker types. The organization of the paper is as follows. Section 2 addresses the related work and inspiration of mixture PLDA model. Section 3 describes the main methods. Section 4 outlines the experimental setup and data preparation.  Section 5 presents the results. Finally, Section 6 concludes this work and mentions future work.  

\section{Related Work}
Child speech has long been studied for automatic speech recognition (ASR) in \protect\cite{potamianos1997automatic,ghai2011study}. Early development of diarization system on children's speech has focused on child language acquisition analysis \protect\cite{najafian2016speaker} or proprietary smart home devices \protect\cite{xu2009reliability,ford2008lenatm}. The diarization of four classes among a primary child, secondary child, adult, and non-speech was first explored in \protect\cite{najafian2016speaker}. The speaker-independent DNN-HMM system \protect\cite{najafian2016speaker} achieved around 20\% DER per child category. However, disentangling the non-speech events and children's speech remains a challenge, as about 10\% of the true child speech was misclassified to be non-speech. One recent work in \protect\cite{sun2018novel} studied the speech enhancement of the noisy environment in daylong realistic recordings, including the SEEDLingS \protect\cite{vandam2016homebank}, a child-centered dataset. The proposed LSTM-based enhancement preprocessor  with a built-in diarization system achieved a 39.2\% DER \protect\cite{vijayasenan2012diartk}. 
The mixture PLDA model has been mainly studied for speaker verification tasks \protect\cite{mak2015mixture,senoussaoui2011mixture}. The work of \protect\cite{senoussaoui2011mixture} showed a better performance using the mixture of two gender-dependent PLDAs than a single gender-independent PLDA. Our work extend from \protect\cite{senoussaoui2011mixture} to three classes of speakers and further develops the mixture of PLDAs for diarization.



\section{Methods}
In this section, we illustrate the methods to incorporate speaker type information to diarization. Subsection 3.1. explains an ideal segmentation step that takes account of oracle speaker type labels. Subsection 3.2. explains the concept of the mixture of PLDA models that encompasses speaker type priors. Lastly, subsection 3.3 describes an estimator of speaker type confidence scores. 
\subsection{Speaker Type Segmentation}
\begin{figure}[ht]
\centering
\includegraphics[scale=0.26]{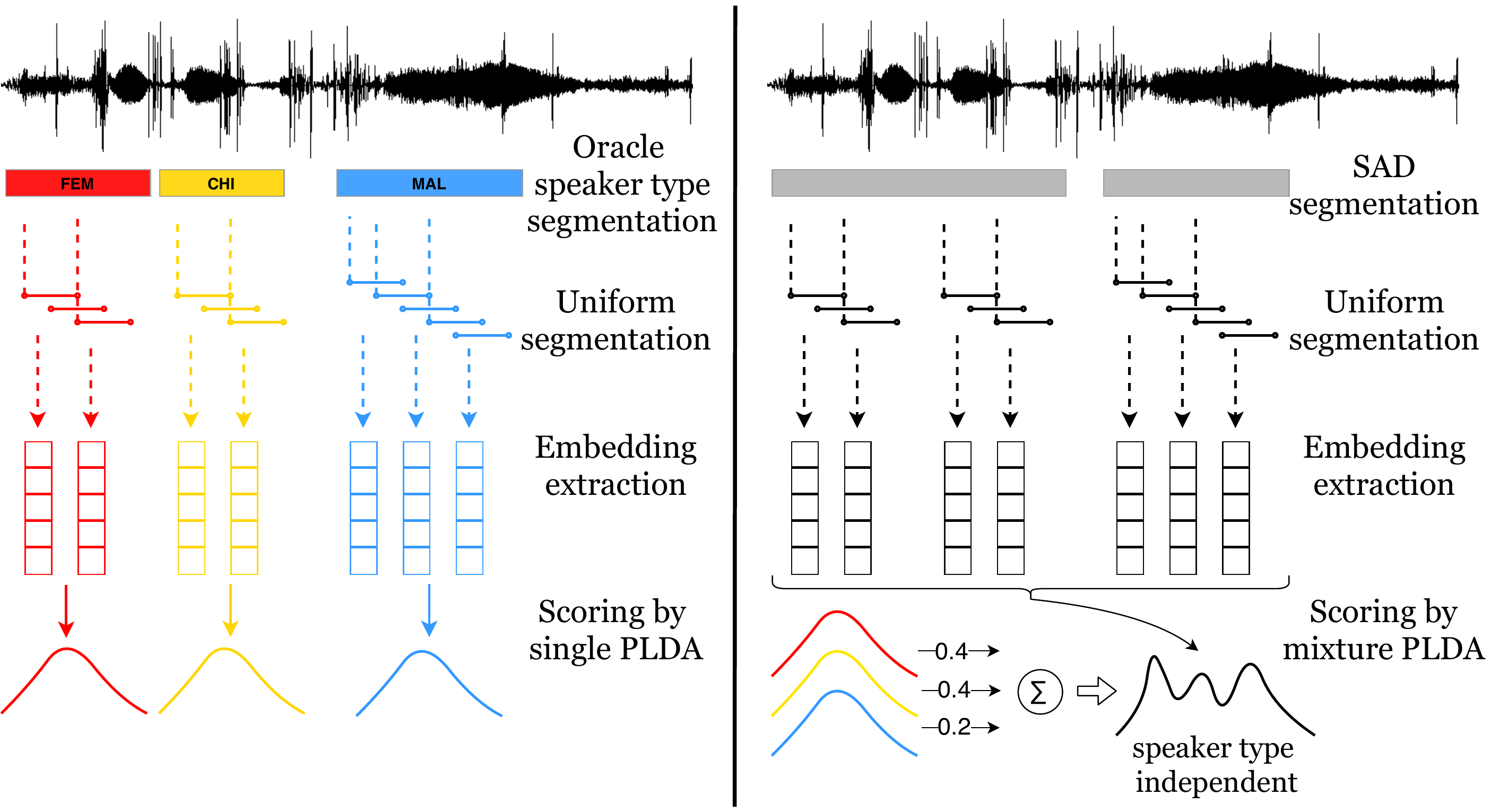}
\caption[]{Diarization steps using speaker type information. Left: oracle speaker type segmentation, Right: mixture PLDA}
\label{fig:oraclesteps}
\end{figure}

The speaker type segmentation refers to the process which splits speech into three parts that each belongs to an adult female, adult male, or child class. Diarization is subsequently performed on each speech region of a class. As illustrated on the left of figure \ref{fig:oraclesteps}, each of the speech splits then goes through the uniform segmentation and scored by a PLDA model trained on the data from the corresponding speaker type. Since the speaker types are mutually exclusive, the diarization proposals of speakers in each of the speech splits will be differentiated, i.e. the \textit{speaker1} of female speech is not the \textit{speaker1} of male speech.
This is of course an ideal setup if provided with the gold speaker type labels. We find that the empirical performance using predicted labels from a classifier is worse compared to a standard baseline because the confusion made early between speaker types can cause the wrong assignments of speakers later.
\subsection{Speaker Type Informed Scoring}
To prevent the hard assignment of speaker types, a probabilistic framework is considered. The similarity scoring in diarization \protect\cite{sell2014speaker} relies on the likelihood ratio between the same-speaker hypothesis $H_s$ and different-speaker hypothesis $H_d$,
\begin{equation}
    \mathcal{R} = P(z_{1},z_{2}|H_{s})/P(z_{1},z_{2}|H_{d})
\end{equation}
where $z_1$ and $z_2$ are the segment-level speaker representations, and the likelihoods may be obtained from either a single PLDA model or a mixture of PLDA models.
\subsubsection{Single PLDA model}
The PLDA model was originally proposed in \protect\cite{ioffe2006probabilistic}, where data variations are captured by a latent class variable. In diarization, speech utterances are thought to vary between and within speakers. The PLDA model is often used to represent the speaker as a latent class variable. The model learns a projection space where distance between representations of different speakers is maximized, and distance between representations of the same speaker is minimized. Given a single PLDA model with the learned covariance $\psi_{s}$ in the transformed space, the likelihood ratio in equation (1) can be represented by,
\begin{align}
    LR(z_1, z_2 | \psi_{s}) =  P(z_{1}|z_{2},\psi_{s})/P(z_{1}|\psi_{s})
\end{align}
where $z_{1}$ and $z_{2}$ are conditionally independent given $H_{d}$. Although the single PLDA model provides a unified framework to compare speakers, it does not explicitly model the known variations of speakers such as gender or age that is often provided as metadata in a dataset.

\subsubsection{Mixture PLDA model based on Speaker Types}
The mixture PLDA model is a formulation that can be thought as a linear combination of different single PLDA models by the weight of a prior. The prior acts as the most general view of data variations. Therefore, we can use the prior of speaker types to weigh each single PLDA model trained on the data from each speaker type. Compared to a single model trained on the data of all speaker types, a mixture PLDA potentially allows an informed discrimination between speakers through the prior. This is figuratively shown on the right of figure \ref{fig:oraclesteps}. 

Under a mixture PLDA model, the numerator of equation (1) can be written as a convex combination of speaker-type dependent PLDA models,
\begin{align}
 P(z_1, z_2 | H_s )&= \sum\limits_{g_1 \in G, g_2 \in G} P(g_1, g_2 | H_s) P(z_1, z_2 | g_1, g_2, H_s)
\end{align}
where $g_1$ and $g_2$ are the speaker types in $G=\{'M', 'F', 'C'\}$ corresponding to $z_1$ and $z_2$, and $'M'$, $'F'$, $'C'$ are the adult male, adult female, and child speaker types, respectively. The denominator of equation (1) can be written as 
\begin{align}
P(z_1, z_2 | H_d) &= \sum\limits_{g_1\in G, g_2 \in G} P(g_1,g_2|H_d) P(z_1, z_2 | g_1, g_2, H_d)
\end{align}
where there are 9 terms in the denominator (all pairwise combinations of speaker types). Given the same speaker, both $g_1$ and $g_2$ must belong to the same speaker type, 

\begin{align*}
    P(g_1, g_2 | H_s)&=P(g_{1}) \\
                    &= P(g_{2}) \\
    P(z_1, z_2 | g_1, g_2, H_s) &= P(z_1, z_2 |  \psi_{g_1}) \\
    &= P(z_1, z_2 |  \psi_{g_2})
\end{align*}

Under different speakers, 
\begin{align*}
    P(g_1, g_2 |H_d)&=P(g_1) \times P(g_2) \\ 
    P(z_1, z_2 | g_1, g_2, H_d) &= P(z_1| \psi _{g_1}) \times P(z_2 | \psi_{g_2})
\end{align*}
where $\psi_{g_1}$ and ${\psi_{g_2}}$ are parameters of the $g_1$-type PLDA and $g_2$-type PLDA, respectively. Here, we assumed the distribution of a speaker type is independent under the different-speaker condition $H_{d}$. The single likelihood $P(z_{i}| \psi _{g_{i}})$ or the joint likelihood $P(z_{1},z_{2}| \psi_{g_{1}})$ can be obtained from the single PLDA model as described in \protect\cite{ioffe2006probabilistic}. But the prior distribution $P(g)$ is a design choice to make. 



\subsection{Speaker Type Confidence Estimator}
As explained in the previous section, the prior distribution of a speaker type is key to the mixture PLDA formulation. One simple way is to assume a constant prior for all recordings encountered in the evaluation. For instance, the prior of child speakers should be above the uniform threshold of 0.33 for diarization on children's speech. We illustrate briefly the other approach to estimate the speaker type confidence from frame-level features. 
\subsubsection{Problem Formulation}
Our goal is to obtain an informed prior probability $P(g)$ for the mixture PLDA. We can consider to use the posterior $P(g|X)$ given an input feature sequence $X$. The confidence estimates for each speaker type can be adopted by taking the softmax output of a neural network \protect\cite{dai2019transformer,sutskever2014sequence}. The prior distribution $P(g_{i})$ that a speech segment $z_{i}$ belongs to a speaker type $g$ for $\forall i \in \{1,2\}$ can be approximated by,
\begin{align}
   P(g_{i}) \approx P_{nn}(g_{i}|z_{i}) = \frac{1}{T_{i}}\ \sum_{t=0}^{T_{i}}\ P_{nn}(g_{i}^{t}|X)
\end{align}
where $z_{i}$ has $T_{i}$ frames and $P_{nn}(g_{i}^{t}|X)$ is the frame-wise posterior output of the network given the whole input sequence. We experimented with this using various sequence-to-sequence and Transformer architectures \protect\cite{dai2019transformer,sutskever2014sequence}, but found that although such a trained system can predict the correct speaker type label with around 75\% accuracy, the performance gains are not transferred when used as mixture PLDA weights, motivating future work on calibration of neural network output probabilities.





\section{Experimental Setup}
The experimental setup is illustrated in this section.  
\subsection{System Description}
Our diarization system mainly follows the x-vector-based system from the DiHARD 2018 \protect\cite{sell2018diarization} recipe in Kaldi \protect\cite{povey2011kaldi}. We focus on extending the PLDA model within this pipeline. 
The audio data input is sampled at 16k-Hz. Mel-frequency cepstral coefficients (MFCC) are used as features and 30 cepstral coefficients are taken from 30 mel-frequency bins. The features are extracted over a 25ms window with a frame rate of 10ms. Both the Delta and the Delta-Delta features are appended. Cepstral mean normalization is applied over a sliding window up to 3 seconds. After the pre-processing, each segment in the {\em evaluation} and the {\em PLDA training} data is subsegmented by a 1.5s sliding window with a 0.75s overlap. The speaker features are then extracted from the subsegments and length normalized \protect\cite{garcia2011analysis}. The x-vector embedding has 512 dimensions.

\subsection{Datasets}
\subsubsection{Adult speech}
We use the Voxceleb \protect\cite{Nagrani17,Chung18b} datasets for the adult speech. The \textit{VoxCeleb1} and \textit{VoxCeleb2} \protect\cite{Nagrani17,Chung18b} are two versions of a large scale dataset that contains interview videos of celebrities uploaded to YouTube. The speakers in the dataset are expected to be mainly adults. There are a total of 7325 speakers with 61\% being male and 39\% being female. We filter by gender in each dataset to train individual PLDA models of adult speaker types.
\subsubsection{Child speech} 
The \textit{CMU Kids} \protect\cite{eskenazi1997cmu} and the \textit{CSLU Kids} corpus \protect\cite{shobaki2007cslu} are used for child speech training. Both datasets were collected for speech recognition tasks, so the audio quality is considered clean. The age of children from both datasets cover a range from 5 to 10 years old. The combined set contains 1191 speakers and about 42.6 hours of speech. The average duration of an utterance is about 4 seconds long.
\subsubsection{Baby vocalization}
We use a subset of the data provided in the Interspeech ComParE challenge \protect\cite{schuller2018interspeech} as the \textit{augmentation} dataset, which collects mostly baby crying sounds. The dataset contains 5.6k recordings with about 2.8 hours of baby vocalizations. We highlight that the average duration of a recording is only 1.8 seconds long, which makes this small dataset hardly sufficient to train models on baby speakers alone.
\subsubsection{BabyTrain test}
The \textit{babyTrain} is a newly aggregated dataset of 9 child-centered corpus \protect\cite{vandam2016homebank} with daylong recordings in the home environment. The \textit{train}, \textit{dev}, and \textit{test} split of the dataset is prepared by the JSALT 2019 workshop \protect\cite{garcia2019speaker} and covers a total of 270 hour recordings. The age of children in the dataset varies between 5 to 60 months old. We adopt the provided dataset splits of 57.5\% \textit{train}, 27\% \textit{development}, and 15.5\% \textit{test} set of the total audio length. The oracle mapping of speakers to categories of key child, child, adult female, adult male, and others is provided. The distribution of speaker type in \textit{train} is similar to the \textit{test}, which about 46\% is child, 50\% is female, and 4\% is male. We exclude recordings where the distant speakers are annotated but with an undefined speaker type label. This leaves us with 329 out of 413 files.

\subsection{Baseline Setup}
Our baseline system uses the single PLDA model trained on the \textit{VoxCeleb1}, and \textit{CMU Kids} and \textit{CSLU Kids} corpus.

\subsection{Oracle Speaker Type Experiment}
To obtain an estimate of the upper bound on the performance of the speaker type informed diarization system, we evaluate the system based on the oracle speaker type labels, that is $p(F)=1$ when the speaker for the speech segment $z$ is Female. This effectively shifts the responsibility of likelihood ratio scoring to one PLDA model trained on the true speaker type.

\subsection{Mixture of Speaker-type PLDAs}
The mixture PLDA is compared to the single PLDA model in the evaluation. Both models are trained on the same dataset, where we further split the data by speaker types to train the mixture PLDA. To study the influence of a data imbalance, we either use the whole dataset or randomly select 1000 speakers from each speaker type to compose the mixture PLDA. We further compare between a nonuniform and uniform prior of speaker types, where the nonuniform distribution on female, child, and male is 40\%, 40\%, and 20\%, respectively.

\subsection{Evaluation Metric}
The primary evaluation metric of our experiments is the diarization error rate (DER). We score speech overlaps and do not use non-score collar. The DER measures a cumulative duration of the following three types of errors over a total duration of valid scoring regions,
\begin{enumerate}
    \item False alarm (FA) – classifying non-speech as speech 
    \item Miss (MS) – classifying speech as non-speech
    \item Speaker mismatch (SM) – actual speaker differs from the claimed speaker
\end{enumerate}
\section{Results}

We conduct three main experiments. The first one examines the upper bound of performance gain using the oracle speaker type segmentation. The second one studies the effectiveness of baby-vocalization augmentation. The last one evaluates the single (UniPLDA) and mixture (MixPLDA) model as well as the influence of training data balance. The system performance is evaluated using DER under the \textit{gold number of speakers} and \textit{oracle speech activity detection}. The main results from the three experiments are presented in Table 1, Table 2, and Table 3. Details of the experiments were illustrated in section 4.
\subsection{Performance Upper Bound on using Speaker Types}
To study the benefit of speaker type information, the evaluation recording is split into three speech regions of each speaker type, using the oracle label. We instead had to use a score threshold to stop the clustering since the number of speakers in each speaker-type segmented audio is unknown.
\begin{table}[ht]
\centering
\renewcommand\arraystretch{1.1}
\begin{tabular}{ccc}
\toprule
UniPLDA Baseln & Oracle Speaker Type & Same Speaker \\
\hline
39.90 (-0.2) & \textbf{28.20} (0.0) & 40.26 (-)\\
\bottomrule
\end{tabular}
\caption{DER(\%) (threshold) comparison between baseline and oracle speaker type}
\label{tab:oss}
\end{table}
 Shown in Table \ref{tab:oss}, using the oracle speaker type reduces the UniPLDA baseline by a significant 11.7\% DER. This verifies our claim that extra speaker type information is beneficial for diarization. The last entry shown in Table \ref{tab:oss} illustrates the worst scenario when the system outputs only one speaker. This result also implies the dominant speaker accounts for about 60\% of the speech (\texttildelow 40\% DER), and of the remaining 40\% belongs to other speakers.

\subsection{Baby Vocalization Augmentation}
The baby vocalization augmentation is found to be an effective domain adaptation of the child-type PLDA. We apply different augmentation techniques to the clean children's speech and compared the results in Table \ref{tab:voc}. 
\begin{table}[th]
\centering
\setlength{\tabcolsep}{6.5pt}
\renewcommand\arraystretch{1.2}
\begin{tabular}{cccccc}
\toprule
System & Clean & mn & v & vn & vmus \\
\hline
CHI-PLDA & 39.61 & 39.28 & 38.74 & \textbf{37.70} & 39.51 \\
\bottomrule
\end{tabular}
\caption{DER(\%) comparison between clean and different augmentations of the child-type PLDA. The music, noise, vocalization, and MUSAN augmentation are labeled accordingly as m,n,v, and mus.}
\label{tab:voc}
\end{table}

As shown from above, using the vocalization (column 4) alone is found to reduce 0.9\% DER from the clean baseline. Compared to the gains from adding double-sized or triple-sized samples generated by music and noise augmentation (column 3) or the MUSAN augmentation (column 6), the vocalization seems to provide the most matched information to the target domain. Lastly, we find the vocalization is complementary with the noise augmentation, which further reduces the clean baseline by an absolute 1.9\% DER.
\subsection{Mixture PLDA and Data Balance}
We observe combining the kids data and \textit{VoxCeleb1} to train the UniPLDA baseline achieves a 38.62\% DER. Our proposed mixture of PLDA models with uniform weights (33\%) on each speaker type, `Male', `Female', and `Child' has a very comparable performance with the UniPLDA. However, with a matched estimation on the speaker type prior to the evaluation, the MixPLDA outperforms the UniPLDA model (row 3 and 5), showing the potential of the speaker type informed model.
\begin{table}[ht]
\centering
\setlength{\tabcolsep}{1.3pt}
\renewcommand\arraystretch{1.2}
\begin{tabular}{ccccc}
\toprule
Training Data & \# utts & UniPLDA & Mix-nunif & Mix-unif \\
\hline
Vox1+Kids & 1.7M & 38.62 & 37.55 & 38.44 \\
Vox12+Kids-vmus & 2.3M & \textbf{35.64} & 37.81 & 37.40 \\
Bal[Vox12+Kids-vmus] & 0.9M & 36.10 & \textbf{35.88} & 36.76 \\
Vox12+Kids-vn & 1.8M & 36.89 & 37.77 & 37.45 \\
Bal[Vox12+Kids-vn] & 0.9M & 37.14 & \textbf{35.83} & 37.18 \\
\bottomrule
\end{tabular}
\caption{DER(\%) comparison between the single (Uni) and mixture PLDA (Mix-nunif and -unif). The size of training data is shown by the number (\#) of Utts. Nunif and unif refers to a nonuniform and a uniform estimate of prior in the mixture PLDA, respectively. Bal[.] indicates that we randomly sample 1k speakers of each speaker type to a balanced dataset. The vocalization, noise and MUSAN\protect\cite{snyder2015musan} augmentation of children data is labeled accordingly as v,n,and mus.}
\label{tab:mixvsuni}
\end{table}
The cost of using a mixture model comes close to the single model since training three models in parallel is possible. The value of the constant prior may be elicited from human expert knowledge, but making either a manual inspection from sampling or an intuitive estimate should be sufficient.

We find the importance of keeping the data balanced while training the MixPLDA. Comparing row 2 to 3 or row 4 to 5 in Table \ref{tab:mixvsuni}, we find on average 1.9\% DER improvement in the nonuniform MixPLDA, even though data size is reduced from the balancing operation. On the contrary, the performance of the UniPLDA depends heavily on the size of the data (the largest is the \textit{vmus}, the second goes the \textit{vn}, and baseline is the least). The formulation of \protect\cite{senoussaoui2011mixture} shows the likelihood ratio obtained under the MixPLDA is equivalent to a weighted sum of each likelihood ratio scored obtained from one of the mixed PLDAs. We suspect this may explain why data balancing is helpful since similarly constrained data can limit the modeling space that each PLDA covers.

\section{Conclusion and Future Work}
 In this paper, we presented a diarization framework using the mixture PLDA model that is targeted at children's speech domain. We discovered speaker type information is beneficial and verified a large upper bound of improvement. Empirically, the mixture of speaker-type PLDA models outperforms the single PLDA model when a balanced training data is used. Though the best result is obtained by the single PLDA, the best mixture PLDA system comes close with an absolute 0.2\% difference in DER and a half of the training size required. Using baby vocalizations as additive background noises has shown matching to the age and acoustic condition of children's speech. Our future work is to develop a confidence score estimator of speaker types using neural networks, as illustrated in the paper.
 
\section{Acknowledgements}
The authors would like to thank Jesús Villalba for the constructive discussion on the mixture PLDA formulation.

\bibliographystyle{IEEEtran}

\bibliography{mybib}


\end{document}